\documentclass[aps,prb,twocolumn,superscriptaddress,floatfix]{revtex4-1}

\usepackage{graphicx,graphics}
\usepackage{dcolumn}
\usepackage{amsmath,amssymb,amsfonts}
\usepackage{latexsym,verbatim}
\usepackage{bm}
\usepackage{color}
\usepackage{ulem}
\usepackage[percent]{overpic}
\usepackage[breaklinks=true,colorlinks,citecolor=blue,linkcolor=blue,urlcolor=blue]{hyperref}

\newcommand\T{\rule{0pt}{2.6ex}}       
\newcommand\B{\rule[-1.2ex]{0pt}{0pt}} 

\usepackage{array}

\begin{document}

\title{Current-induced birefringent absorption and non-reciprocal plasmons in graphene}

\author{Ben Van Duppen}
\email{ben.vanduppen@uantwerpen.be}
\affiliation{Department of Physics, University of Antwerp, Groenenborgerlaan 171, B-2020 Antwerp, Belgium}

\author{Andrea Tomadin}
\affiliation{NEST, Istituto Nanoscienze-CNR and Scuola Normale Superiore, I-56126 Pisa,~Italy}

\author{Alexander N. Grigorenko}
\affiliation{School of Physics and Astronomy, University of Manchester, Manchester M13 9PL, UK}

\author{Marco Polini}
\affiliation{Istituto Italiano di Tecnologia, Graphene Labs, Via Morego 30, I-16163 Genova,~Italy}

\begin{abstract}
We present extensive calculations of the optical and plasmonic properties of a graphene sheet carrying a dc current. 
By calculating analytically the density-density response function of current-carrying states at finite temperature, 
we demonstrate that an applied dc current modifies the Pauli blocking mechanism and that absorption acquires a birefringent character with respect to the angle between the in-plane light polarization and current flow. Employing the random phase approximation at finite temperature, we show that graphene plasmons display a degree of non-reciprocity and collimation that can be tuned with the applied current. We discuss the possibility to measure these effects.
\end{abstract}

\maketitle

\section{Introduction}
The photonic, optoelectronic, and plasmonic properties of graphene  
have been attracting a great deal of attention for the past ten years~\cite{bonaccorso_naturephoton_2010,peres_rmp_2010,bao_acsnano_2012,grigorenko_naturephoton_2012,novoselov_nature_2012,koppens_naturenano_2014,basov_rmp_2014,low_acsnano_2014,abajo_acsphoton_2014,roadmap}. This interest was sparked by a series of key experimental results that were demonstrated soon after the isolation of graphene~\cite{geim_naturemater_2007}. Early on, it was shown~\cite{kuzmenko_prl_2008,nair_science_2008} that the optical conductivity of graphene is universal in the limit $\hbar\omega > 2E_{\rm F}$ and given by $\sigma_{\rm uni} = e^2/(4\hbar)$, where $\omega$ is the photon frequency and $E_{\rm F}$ is the Fermi energy. In the same frequency range, absorption is given by $\pi \alpha_{\rm QED}\simeq 2.3\%$, where $\alpha_{\rm QED} = e^2/(\hbar c) \simeq 1/137$ is the quantum electrodynamics fine structure constant. Importantly, absorption can be tuned by gating~\cite{wang_science_2008} and can be enhanced by using a variety of plasmonic protocols~\cite{thongrattanasiri_prl_2012,nikitin_prb_2012,stauber_prl_2014,echtermeyer_naturecommun_2011,liu_naturecommun_2011,freitag_naturecommun_2013} or by embedding graphene in a optical cavity~\cite{engel_naturecommun_2012,furchi_nanolett_2012}. At low photon energies (i.e.~for $\hbar\omega\ll 2E_{\rm F}$), absorption occurs only due to the excitation of intra-band electron-hole pairs, which is allowed by momentum-non-conserving collisions.
 
More recently, Dirac plasmons, the self-sustained density oscillations of the two-dimensional (2D) electron system in a doped graphene sheet, have been studied experimentally by a variety of spectroscopic methods~\cite{grigorenko_naturephoton_2012}. Fei {\it et al.}~\cite{fei_nature_2012} and Chen {\it et al.}~\cite{chen_nature_2012} carried out seminal scattering-type scanning near-field optical microscopy (s-SNOM) experiments in which Dirac plasmons were launched and imaged in real space. They showed that, in the mid infrared, the plasmon wavelength $\lambda_{\rm p}$ can be $\sim40$-$60$ times smaller than the free-space excitation wavelength $\lambda_0 =2\pi c/\omega$, allowing an extreme concentration of electromagnetic energy, and that Dirac plasmon properties are easily gate tunable. These figures of merit have been tremendously improved by utilizing vertical stacks comprising graphene sheets encapsulated in hexagonal boron nitride crystals~\cite{mayorov_nanolett_2011,wang_science_2013}, in the quest for a low-loss and tunable platform for infrared and THz plasmonics. Mid-infrared plasmons in encapsulated graphene sheets display~\cite{woessner_naturemater_2015} ultra-large field confinement (i.e.~$\lambda_{\rm p}/\lambda_0 \sim 1/150$), small group velocity, and a long lifetime exceeding~$500~{\rm fs}$.

In this Article we revisit the optical and plasmonic properties 
of single-layer graphene which carriers a dc current. 
Our interest is motivated by the simple observation that graphene sheets can support ultra-high current densities without being damaged. Indeed, early on Moser~{\it et al.}~\cite{moser_apl_2007} proposed a simple method to suppress contamination of graphene at low temperatures inside a cryostat. Applying a source-drain bias 
of a few Volts across their samples, these authors induced a large current flow of up to $4~{\rm mA}$, 
thereby removing contamination adsorbed on the surface.
For a tipical device width 
of $4~{\rm \mu m}$, such current translates into an extremely large current density 
$J \simeq 10~{\rm A}/{\rm cm}$. 

Here, we present a detailed theoretical study of the optical and plasmonic properties of current-carrying graphene sheets. 
Previous authors~\cite{Vasko2010, Sabbaghi2015} have considered the same problem, 
but only reported approximate solutions that are valid for low current densities. 
Here, we describe a semi-analytical approach, which can be applied for any current density (i.e. for drift velocities $|{\bm v}|$ as large as the graphene Fermi velocity $v_{\rm F}$) and at an arbitrary electron temperature $T$. 
The combined functionalities offered by electrical (or chemical) doping and dc currents may further enrich the graphene optoelectronic and plasmonic portfolio, possibly leading to interesting discoveries and/or potential applications.

Our Article is organized as following. In Sect.~\ref{sect:polarization} we present analytical expressions for the imaginary part of the density-density response function of a current-carrying graphene 
sheet at zero temperature. The calculation of the real part is reduced to a one-dimensional quadrature.
Results at finite $T$ are then obtained by using an integral identity due to Maldague~\cite{Maldague1978,Giuliani_and_Vignale}. In Sect.~\ref{sect:optical} we use the results of Sect.~\ref{sect:polarization} to calculate the absorption as a function of photon frequency $\omega$, drift velocity ${\bm v}$, and temperature $T$. We also calculate analytically the dependence of this quantity on the angle between the polarization of light and current direction. In Sect.~\ref{sect:plasmons} we present our results for the plasmon modes of a current-carrying graphene sheet and quantify the degree of in-plane anisotropy induced by the current flow. In Sect.~\ref{sect:relation} we present mathematical equations relating the chemical potential $\mu$ and drift velocity ${\bm v}$ to the carrier density $n_{\rm c}$ and current density ${\bm J}$. Finally, in Sect.~\ref{sect:summary} we summarize our main findings and discuss potential obstacles against their experimental observation.
\section{The density-density response function of a current-carrying graphene sheet}
\label{sect:polarization}

The optical properties of a current-carrying graphene sheet can be calculated by using linear response theory~\cite{Giuliani_and_Vignale} with respect to the electric field of the incident light beam. We treat the effect of an external dc current in a non-perturbative fashion. To this end, we first find the distribution function of a homogeneous {\it equilibrium} state carrying a finite current, at a finite temperature $T$. The distribution function of such a state can be calculated by requiring that it nullifies the collision integral for electron-electron scattering processes, which conserve energy and momentum~\cite{GantmakherBook}. The end result of this calculation is the following Fermi-Dirac distribution function~\cite{GantmakherBook,bistritzer_prb_2009}:
\begin{equation}\label{eq:FDDistribution}
n_{{\rm F}, \lambda}(\bm{k}, \bm{v}; T) =\left[ \exp \left(\frac{
\varepsilon _{\bm{k},\lambda }-\hbar \bm{v}\cdot \bm{k}-\mu }{k_{\rm B}T}
\right) +1\right]^{-1}~,
\end{equation}
where $\varepsilon _{\bm{k},\lambda}=\lambda \hbar v_{\rm F}|\bm{k}|$ is the Dirac band energy, $v_{\rm F} \sim 10^{6}~{\rm m}/{\rm s}$ is the Fermi velocity, and $\lambda=+$ ($\lambda = -$) denotes conduction (valance) band states. The function $n_{{\rm F}, \lambda}(\bm{k}, \bm{v};T)$ is  parametrized by the drift velocity ${\bm v}$ and chemical potential $\mu$. The drift velocity will be related to the current density ${\bm J} = J {\hat {\bm v}}$ with ${\hat {\bm v}} = {\bm v}/v$ and $v = |{\bm v}|$ below. Since we are dealing with an open system, i.e.~a graphene sheet attached to leads, we do not follow the usual grand-canonical procedure~\cite{ramezanali_jphysA_2009} in which the chemical potential $\mu$ is forced to readjust at each value of $T$ to make sure that the average number of particles is conserved. In this Article, $\mu$ is fixed by the leads (and external gating) and does not change with temperature. (For the sake of definiteness, we consider an $n$-doped graphene sheet: results for $p$-doped systems are identical since the quantities we are interested in are particle-hole symmetric.) 

As we will see below, the drift velocity satisfies the following inequality: $v < v_{\rm F}$. Our results, however, are not perturbative in the dimensionless parameter
\begin{equation}
\beta = \frac{v}{v_{\rm F}}~.
\end{equation}

To find the optical response of a current-carrying two-dimensional (2D) 
massless Dirac fermion (MDF) system~\cite{kotov_rmp_2012}, we calculate the
dynamical density-density response function, which is defined by~\cite{Giuliani_and_Vignale}
\begin{eqnarray}\label{FullResponseFunction}
\chi^{(0)}_{nn}(\bm{q},\omega; T)
&=&N_{\rm f}\sum_{\lambda ,\lambda ^{\prime }}\int \frac{d^2{\bm k}}{(2\pi)^2}F_{\lambda\lambda ^{\prime
}}( \bm{k},\bm{k}+\bm{q})\nonumber\\
&\times&\frac{n_{{\rm F},\lambda}(\bm{k}, \bm{v}; T) -n_{{\rm F}, \lambda'}
(\bm{k}+\bm{q}, \bm{v}; T)}{\hbar \omega +\varepsilon _{\bm{k},\lambda}-\varepsilon _{\bm{k}+\bm{q},\lambda ^{\prime }}+i\eta}~.
\end{eqnarray}
Here, $N_{\rm f} = 4$ is the number of fermion flavors in graphene,
\begin{equation}\label{eq:form_factor}
F_{\lambda\lambda ^{\prime }}(\bm{k},\bm{k}+\bm{q}) = 
\frac{1}{2}\left[ 1+\lambda \lambda ^{\prime }\cos ( \theta _{\bm{k}} - 
\theta _{\bm{k}+\bm{q}})\right]
\end{equation}
is the usual chirality factor~\cite{kotov_rmp_2012}, and $\eta=0^+$ is a positive infinitesimal. 

\subsection{The imaginary part of $\chi^{(0)}_{nn}$ at zero temperature}
\label{sect:imag}

Following standard practice~\cite{Giuliani_and_Vignale}, we first calculate analytically the imaginary part of $\chi^{(0)}_{nn}(\bm{q},\omega; T)$ at $T=0$. 

It is convenient to introduce the following dimensionless wave number
\begin{equation}\label{eq:qbar}
{\bar q}  = \frac{\hbar v_{\rm F} q}{\mu}
\end{equation}
(with $q=|{\bm q}|$) and dimensionless frequency
\begin{equation}\label{eq:omegabar}
{\bar \omega} = \frac{\hbar \omega}{\mu}~.
\end{equation}

After lengthy but straightforward calculations we find the following result:
\begin{equation}\label{Eq:SummaryImaginaryPart}
\text{Im}[\chi^{(0)}_{nn}(\bm{q},\omega; 0)] =f_{\mu}({\bar q}, {\bar \omega}) \sum_{\gamma =\pm 1}U({\bar q}, {\bar \omega}; \beta, \gamma\psi)~,
\end{equation}
where $\psi$ is the angle between ${\bm q}$ and ${\hat {\bm v}}$, i.e.~
\begin{equation}\label{eq:defpsi}
\cos{(\psi)} = \frac{{\bm q}\cdot {\hat {\bm v}}}{q}~,
\end{equation}
\begin{equation}
f_{\mu}(x,y) = \frac{D(\mu) }{16}\frac{x^{2}}{
\sqrt{\left\vert x^{2} - y^{2}\right\vert }}~,
\end{equation}
$D(\mu) = N_{\rm f} \vert \mu\vert/(2\pi \hbar^2 v^2_{\rm F})$ is the 2D MDF density of states, and 
\begin{eqnarray}\label{FFunction}
U({\bar q},{\bar \omega}; \beta, \psi) &=&-\Theta({\bar q}-{\bar \omega}) \sum_{\alpha = \pm 1} \alpha H_{\text{intra}}(A_{\alpha},B,C_{\alpha})  \notag \\
&+&\Theta({\bar \omega} -{\bar q})[H_{\text{inter}}(A_{-1},B,C_{-1})-\pi]~. 
\nonumber\\
\end{eqnarray}
Analytical expressions for the real functions $H_{\text{intra}}(x,y,z)$ and $H_{\text{inter}}(x,y,z)$ in Eq.~(\ref{FFunction}) are reported in Tables~\ref{Tab:Intra} and~\ref{Tab:Inter}, while the quantities $A_{\alpha}$, $B$ and $C_{\alpha}$ are defined as following:
\begin{eqnarray}\label{eq:definitionsABC}
A_{\alpha} &=&2+\alpha {\bar \omega} -\alpha \beta\cos ( \psi ) {\bar q}, \nonumber\\
B &=&\beta{\bar \omega} \cos(\psi) -{\bar q}, \nonumber\\
C_{\alpha} &=&\alpha \beta\sin(\psi)\sqrt{\vert {\bar q}^{2}-{\bar \omega}^{2}\vert }.
\end{eqnarray}

\begin{table*}[tbp]
\begin{tabular}{|c|c|}
\hline
 $A_{\alpha}+B>0$ & $A_{\alpha}+B<0$ \T \B \\ \hline
 $G_{\text{intra}}(
x_{-}^{\alpha }) $ & $\Theta ( A_{\alpha}-\sqrt{B^{2}-C_{\alpha}^{2}}) \left[
G_{\text{intra}}(x_{-}^{\alpha}) -G_{\text{intra}}(x_{+}^{\alpha}) \right]$ \T \B \\ \hline
\end{tabular}
\caption{This table summarizes the expressions that the function $H_{\rm intra}(x,y,z)$ takes depending on the  sign of the quantity $A_{\alpha}+ B$. The left (right) column applies if $A_{\alpha}+ B>0$ ($A_{\alpha}+ B<0$). The function $G_{\text{intra}}( x) = x\sqrt{x^{2}-1}-\log ( x+
\sqrt{x^{2}-1})$ is evaluated at $x = x_{i}^{\alpha}$, where $x_{i}^{\alpha}$ is the root of the function $g_{\text{intra}}(x) =A_{\alpha}+Bx-C_{\alpha}\sqrt{x^{2}-1}$. Here, the subscript $i = \pm$ of the root $x_{i}^{\alpha}$ is given by the sign of the derivative $g^{\prime}_{\text{intra}}(x)$ evaluated at $x = x_{i}^{\alpha}$, i.e.~$g^{\prime}_{\text{intra}}(x_{+}^{\alpha})>0$, while $g^{\prime}_{\text{intra}}(x_{-}^{\alpha})<0$.\label{Tab:Intra}}
\end{table*}

\begin{table*}[tbp]
\begin{tabular}{|c|c|c|}
\hline
 &   $A_{-}+B>0$ &  $A_{-}+B<0$ \\
\hline 
& & \\[-13pt]
$A_{-}-B>0$ &  $
\begin{array}{c}
\Theta\left[ A_{-}-\text{sgn}(C_{-})\sqrt{B^{2}+C_{-}^{2}}\right] \left[G_{\text{inter}}(y_{-})  - 
G_{\text{inter}}(y_{+}) +\pi \right] 
\\ 
+\Theta \left[ \text{sgn}\left[ C_{-}\right] \sqrt{B^{2}+C_{-}^{2}}-A_{-}\right]  
\end{array}$
  & $G_{\text{inter}}( y_{-}) + \pi/2$     \\ 
  & & \\ [-13pt]
\hline
& & \\[-13pt]
$A_{-}-B<0$   & $\pi/2-G_{\text{inter}}(y_{+}) $ & $\Theta 
\left[ A_{-}-\text{sgn}(C_{-})\sqrt{B^{2}+C_{-}^{2}}\right] \left[ G_{\text{
inter}}(y_{-}) -G_{\text{inter}}(y_{+}) \right] $  \\ [3pt]
\hline
\end{tabular}
\caption{Same as in Table~\ref{Tab:Intra}, but for the function $H_{\rm inter}(x,y,z)$. Here, $G_{\text{inter}}(x) = x\sqrt{1-x^{2}}+\arcsin(x)$ and $y_{i}$ with $i = \pm 1$ denote the roots of the function $g_{\text{inter}}(y) =A_{-}+By-C_{-}\sqrt{1-y^{2}}$. As in Table~\ref{Tab:Intra}, the subscript $i$ is such that 
$g^{\prime}_{\text{inter}}(y_{+})>0$, while $g^{\prime}_{\text{inter}}(y_{-})<0$.\label{Tab:Inter}}
\end{table*}

\subsection{The real part of $\chi^{(0)}_{nn}$ at zero temperature}
\label{sect:real}

The real part of $\chi^{(0)}_{nn}(\bm{q},\omega; T)$ at $T = 0$ can be calculated from the following integral
\begin{eqnarray}\label{PrincipalValue}
{\rm Re}[\chi^{(0)}_{nn}(\bm{q},\omega; 0)]
&=&N_{\rm f}\sum_{\lambda ,\lambda ^{\prime }}{\cal P}\int \frac{d^2{\bm k}}{(2\pi)^2}F_{\lambda\lambda ^{\prime
}}( \bm{k},\bm{k}+\bm{q})\nonumber\\
&\times&\frac{n_{{\rm F},\lambda}(\bm{k}, \bm{v}; 0) -n_{{\rm F}, \lambda'}
(\bm{k}+\bm{q}, \bm{v}; 0)}{\hbar \omega +\varepsilon _{\bm{k},\lambda}-\varepsilon _{\bm{k}+\bm{q},\lambda ^{\prime }}}~,\nonumber\\
\end{eqnarray}
where ${\cal P}$ denotes a principal-value integration. 

Without loss of generality, we can take ${\bm q} = q {\hat {\bm x}}$ and denote by $\theta$ the polar angle of ${\bm k}$. We first compute {\it analytically}  the integral over $k = |{\bm k}|$ in Eq.~(\ref{PrincipalValue}) and then treat the angular integration over $\theta$ numerically. By combining terms from the sum over $\lambda$ and $\lambda^{\prime}$, the integral in Eq.~(\ref{PrincipalValue}) can be written as the sum of two contributions,
\begin{equation}
{\rm Re}[\chi^{(0)}_{nn}(\bm{q},\omega; 0)] = {\rm Re}[\delta \chi^{(0)}_{nn}(\bm{q},\omega; 0)] + {\rm Re}[\chi_{\rm u}(\bm{q},\omega)]~,
\end{equation}
where
\begin{equation}
\text{Re}[\chi_{\rm u} (\bm{q},\omega)] = -\frac{ N_{\rm f} q^{2} \Theta(v_{\rm F}q-\hbar \omega)}{16 \sqrt{(\hbar v_{\rm F}q)^{2}-(\hbar\omega)^{2}}}~,
\end{equation}
is the undoped (or vacuum) contribution~\cite{wunsch_njp_2006} and
\begin{eqnarray}\label{eq:inter_and_intra_parts}
{\rm Re}[\delta \chi^{(0)}_{nn}(\bm{q},\omega; 0)] &=& N_{\rm f}\sum_{\alpha, \alpha^{\prime} = \pm 1}{\cal P}\int \frac{d^2{\bm k}}{(2\pi)^2}F_{\alpha^{\prime}}( \bm{k},\bm{k}+\bm{q})\nonumber\\&\times & \frac{\alpha n_{\rm F}(\bm{k}, \alpha \bm{v}; 0)}{\hbar \omega +\alpha(\varepsilon _{\bm{k}}- \alpha^{\prime} \varepsilon_{\bm{k}+\bm{q}})}
\end{eqnarray}
is the contribution due to a finite doping. In Eq.~(\ref{eq:inter_and_intra_parts}) the distribution functions and band energies are evaluated for $\lambda = +$, i.e. $\varepsilon _{\bm k} = \varepsilon _{\bm{k},+}$ and $n_{\rm F}({\bm k}, {\bm v}; 0) = n_{{\rm F},+}({\bm k}, {\bm v}; 0)$.

To calculate the integral in Eq.~(\ref{eq:inter_and_intra_parts}) we observe that, in the absence of a drift velocity ($\beta = 0$), the upper limit of the integral over $k$ is the Fermi wave number $k_{\rm F} = \mu/(\hbar v_{\rm F})$. Indeed, for $\beta=0$ the Fermi-Dirac distribution function in Eq.~(\ref{eq:inter_and_intra_parts}) has the form:
\begin{equation}\label{eq:fermi_dirac_no_drift}
n_{{\rm F}}({\bm k},0;0)=\Theta(\mu - \hbar v_{\rm F} k)= \Theta(k_{\rm F}- k)~,
\end{equation}
where $\Theta(x)$ is the Heaviside step function. 

For $\beta>0$ the distribution function changes into 
\begin{eqnarray}\label{eq:fermi_direc_finite_drift}
n_{{\rm F}}({\bm k}, \alpha{\bm v};0) &=& \Theta(\mu - \hbar v_{\rm F} k (1-\alpha \beta \cos(\psi - \theta)))~,\nonumber\\
&=& \Theta(k_{\rm F}(\theta)- k)~,
\end{eqnarray} 
where, in analogy with Eq.~(\ref{eq:fermi_dirac_no_drift}), we have introduced an angle-dependent Fermi wave number:
\begin{equation}
k_{\rm F}(\theta) = \frac{\mu}{\hbar v_{\rm F}[1-\alpha \beta \cos(\psi - \theta)]}~.
\end{equation}
This allows us to rewrite the real part of the response function as
\begin{equation}\label{Eq:RealIntegral}
{\rm Re}[\delta \chi^{(0)}_{nn}(\bm{q},\omega; 0)] = D(\mu) \sum_{\alpha = \pm 1}  \int_{0}^{2\pi} \frac{d\theta}{2\pi }\frac{I(\theta) }{ \alpha {\bar \omega} -{\bar q}\cos \theta}~,
\end{equation}
where
\begin{equation}\label{eq:principal_value_integral}
I(\theta) = {\cal P}\int_{0}^{\bar{\kappa}_{\rm F}(\theta) }dkk\frac{k+X(\theta)}{k+Y(\theta)}~.
\end{equation}
In Eq.~(\ref{eq:principal_value_integral}) we have defined:
\begin{equation}
\bar{\kappa}_{\rm F}(\theta) = \frac{\hbar v_{\rm F}k_{\rm F}(\theta)}{\mu}~,
\end{equation}
\begin{equation}
X(\theta)=\frac{\alpha {\bar \omega} +{\bar q}\cos \theta }{2}
\end{equation}
and
\begin{equation}
Y(\theta) =\frac{{\bar \omega}^{2}-{\bar q}^{2}}{2( \alpha {\bar \omega} -{\bar q}\cos \theta ) }~.
\end{equation}
The quadrature in Eq.~(\ref{eq:principal_value_integral}) can be carried out analytically. The end result is
\begin{eqnarray}
I&=&\frac{1}{2}\bar{\kappa}_{\rm F}( 2X-2Y-\bar{\kappa}_{\rm F})\\ && + Y(X-Y) \log \left\vert \frac{Y}{Y+\bar{\kappa}_{\rm F}}
\right\vert~.
\end{eqnarray}
Having found an expression for $I(\theta)$, the angular integration in Eq.~(\ref{Eq:RealIntegral}) can be easily carried out numerically.

\begin{figure}[tb]
\centering
\includegraphics[width= 8.5cm]{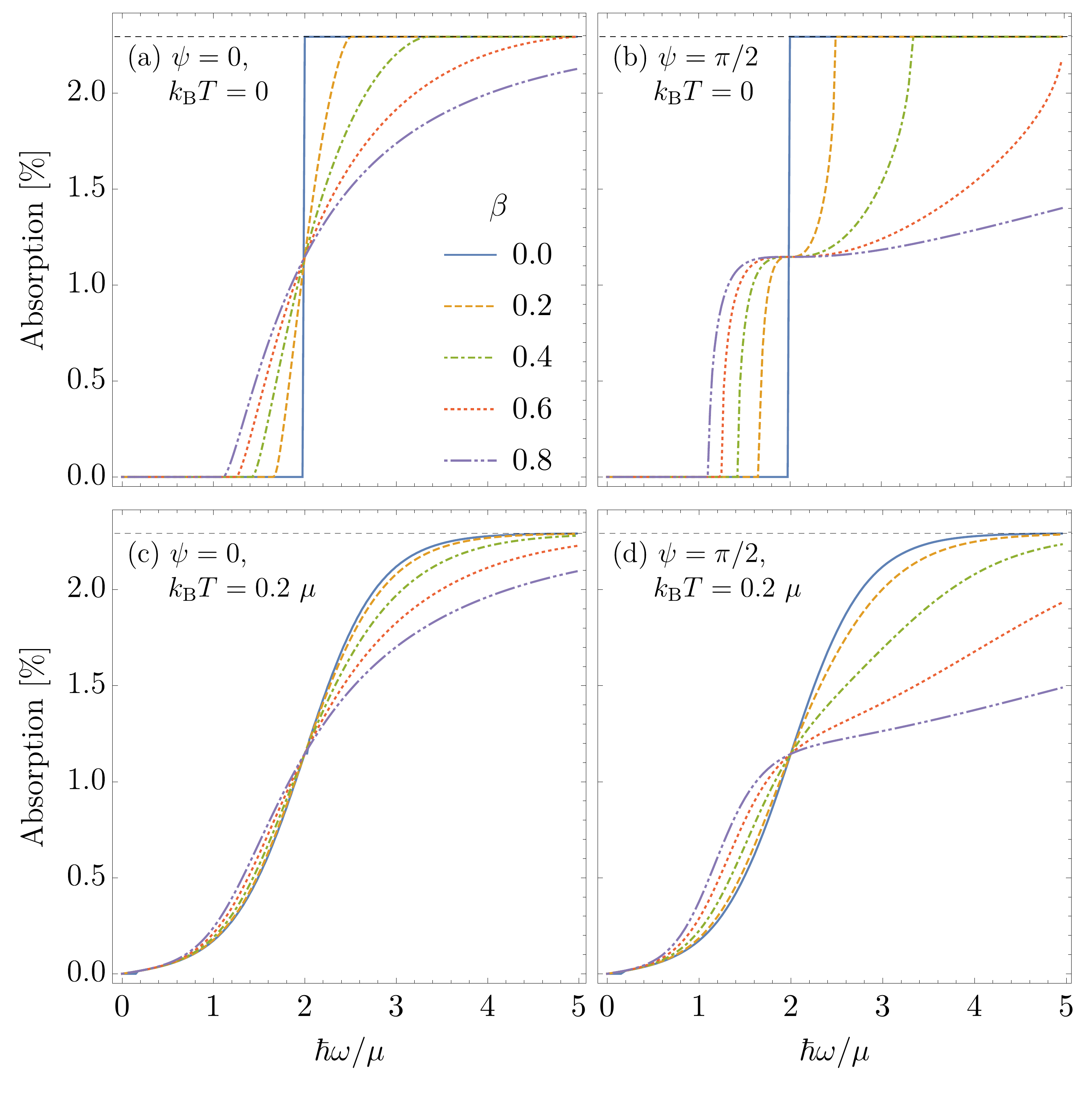}
\caption{(Color online) Absorption as a function of the incident photon energy (in units of $\mu/\hbar$), for different values of the dimensionless drift velocity $\beta=v/v_{\rm F}$. Panels (a) and (b) show zero-temperature results for a polarization of light along ($\psi=0$) and perpendicular ($\psi=\pi/2$) to the velocity direction, respectively. Panels (c) and (d) show the same quantities but for a finite temperature $k_{\rm B} T = 0.2~\mu$. The dashed horizontal curve in each panel indicates graphene's maximal absorption, $\pi \alpha_{\rm QED} \simeq 2.3\%$. \label{Fig:AbsCurves}}
\end{figure}
\subsection{Extension to finite temperature}
\label{sect:finiteT}

The temperature dependence of the response function in Eq.~$( \ref
{FullResponseFunction})$ is best evaluated by using the following elegant identity due to 
Maldague~\cite{Maldague1978, Giuliani_and_Vignale}:
\begin{equation}\label{eq:maldague_mathematical_identity}
\frac{1}{\exp(x)+1} = \int_{-\infty}^{\infty} dy \frac{\Theta(y-x)}{4\cosh^{2}(y/2)}~.
\end{equation}
Because the Fermi-Dirac distribution at $T=0$---see Eq.~(\ref{eq:fermi_direc_finite_drift})---has the form of a Heaviside step function, we can use Eq.~(\ref{eq:maldague_mathematical_identity}) to write the following integral representation:
\begin{equation}\label{eq:maldague}
n_{{\rm F},\lambda}({\bm k}, {\bm v}; T) =\int_{-\infty }^{\infty }dy
\frac{[n_{{\rm F},\lambda}({\bm k}, {\bm v}; 0)]_{\mu \to y}}{4k_{\rm B}T \cosh^{2}\left[
(y - \mu)/(2k_{\rm B}T)\right]}~.
\end{equation}
In Eq.~(\ref{eq:maldague}) $n_{{\rm F},\lambda}({\bm k}, {\bm v}; T)$ is expressed as an integral over $\mu$ of the zero-temperature Fermi step, accompanied by an appropriate weighting function. We then observe that the difference between distribution functions in the numerator of Eq.~(\ref{FullResponseFunction}) can be rewritten as:
\begin{equation}\label{eq:difference_distribution_function}
\int_{-\infty}^{\infty}dy\frac{\left[n_{{\rm F},\lambda}(\bm{k}, \bm{v}; 0)\right]_{\mu \to y} - \left[n_{{\rm F}, \lambda'}(\bm{k}+\bm{q}, \bm{v}; 0)\right]_{\mu \to y}}{4 k_{\rm B} T \cosh ^{2}[(y-\mu)/(2k_{\rm B}T)]}~.
\end{equation}
Since $\mu$ appears only in the numerator of Eq.~(\ref{FullResponseFunction}), we can integrate first over ${\bm k}$ and {\it then} over $y$ obtaining:
\begin{equation}\label{eq:finite_temp_chi}
\chi_{nn}^{( 0) }( \bm{q},\omega ;T) = \int_{-\infty }^{\infty }dy \frac{ \left[\chi_{nn}^{( 0) }( \bm{q},\omega ;0)\right]_{\mu \to y}}{4 k_{\rm B}T \cosh^{2}[(y-\mu)/(2k_{\rm B}T)]}~.
\end{equation}
This formula expresses the finite temperature response function $\chi_{nn}^{(0)}({\bm q}, \omega; T)$  in terms of a weighted average of  response functions  $\chi_{nn}^{( 0) }( \bm{q},\omega ;0)|_{\mu \to y}$ at zero temperature and different chemical potentials.

\begin{figure}[t]
\centering
\includegraphics[width= 9cm]{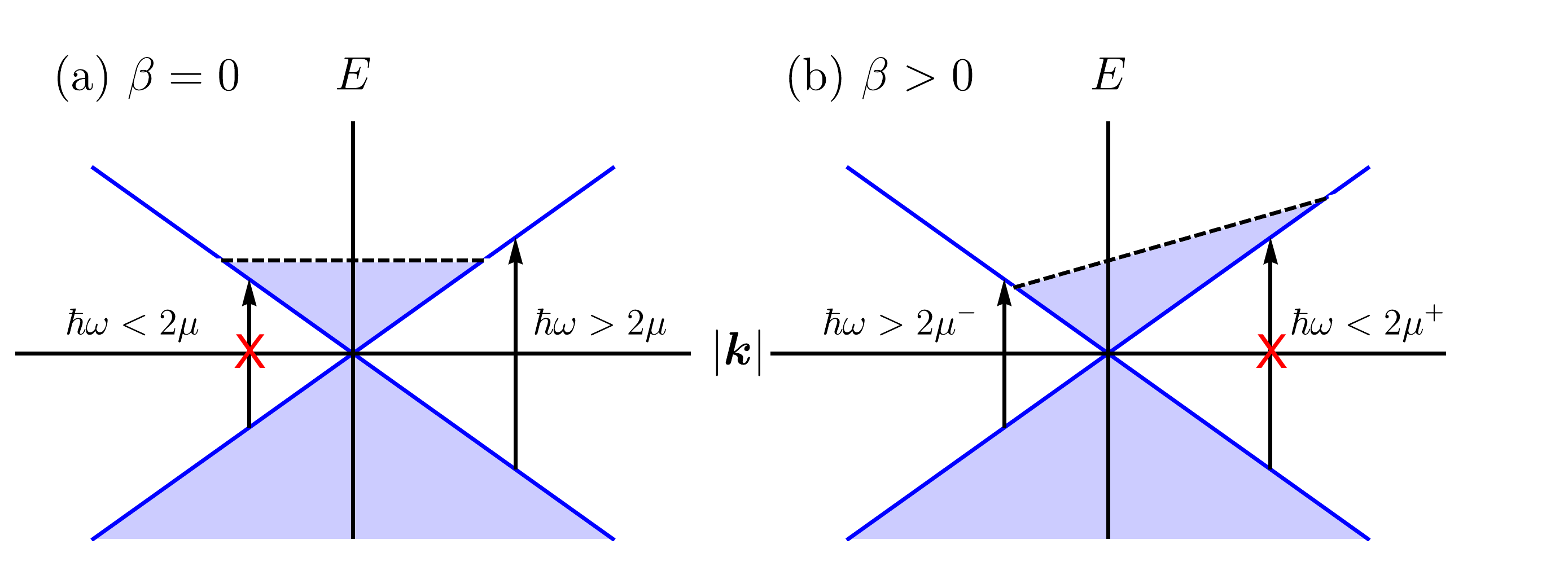}
\caption{(Color online) Illustration of the effect of a finite drift velocity on the Pauli
blocking mechanism. Vertical arrows indicate electron-hole pair
creation, while the red cross indicates transitions that are forbidden due to Pauli blocking. 
Panel (a) shows the Pauli blocking mechanism in the case $\beta =0$, while panel (b) pictures the effect of a tilted Fermi level due to a finite value of $\beta$. The quantities $\mu^{-}$ ($<\mu$) and $\mu^+$ ($>\mu$) have been defined in Eq.~(\ref{eq:muplusminus}). The sharp absorption step $\pi\alpha_{\rm QED}\Theta(\hbar \omega- 2\mu)$ at $\beta =0$ is smoothened by a finite value of $\beta$ because Pauli blocking: 
i) is suppressed in the window $2\mu^-<\hbar \omega < 2\mu$ and ii) extends its action {\it above} $2\mu$ in the window $2\mu<\hbar \omega<2\mu^+$.\label{Fig:Sketch}}
\end{figure}
\section{Birefringent absorption}
\label{sect:optical}

In this Section we use the results for the density-density response function reported in Sect.~\ref{sect:polarization} to calculate how the optical absorption spectrum of graphene is changed by a dc current. 

Because the drift velocity ${\bm v} = (v_x,v_y)$ defines a preferential direction in the graphene sheet, the conductivity 
$\sigma_{ij}(\omega)$ stemming from the density-density response function 
$\chi_{nn}^{(0)}({\bm q},\omega ;T)$ in Eq.~(\ref{eq:finite_temp_chi}) transforms like a rank-$2$ tensor with respect to the Cartesian indices $i,j = x,y$.  Following common practice---see Section~3.4 in Ref.~\onlinecite{Giuliani_and_Vignale}---in the case of similar rank-$2$ tensors (like the current-current response function of a homogeneous and isotropic electron system), we can decompose the tensor $\sigma_{ij}(\omega)$ according to
\begin{equation}\label{eq:rank-two-tensor}
\sigma_{ij}(\omega) = \sigma_{\rm L}(\omega)\frac{v_{i}v_{j}}{v^2} + 
\sigma_{\rm T}(\omega)\left(\delta_{ij} - \frac{v_{i}v_{j}}{v^2}\right)~,
\end{equation}
where the longitudinal $\sigma_{\rm L}(\omega)$ and transverse $\sigma_{\rm T}(\omega)$ components are defined by:
\begin{equation}\label{eq:optical_conductivity_longitudinal}
\sigma_{\rm L}(\omega) \equiv i e^2 \omega \left.\lim_{q\to 0}\frac{\chi^{(0)}_{nn}({\bm q}, \omega; T)}{q^2}\right|_{\psi = 0}
\end{equation}
and
\begin{equation}\label{eq:optical_conductivity_transverse}
\sigma_{\rm T}(\omega) \equiv i e^2 \omega \left.\lim_{q\to 0}\frac{\chi^{(0)}_{nn}({\bm q}, \omega; T)}{q^2}\right|_{\psi = \pi/2}~.
\end{equation}
Of course, $\sigma_{\rm L}(\omega)$ and $\sigma_{\rm T}(\omega)$ depend on $v$.

The optical absorption $A_{ij}(\omega)$ is defined by
\begin{equation} \label{eq:optical-absorption}
A_{ij}(\omega) \equiv \frac{4\pi}{c}\text{Re}[\sigma_{ij}(\omega)]~.
\end{equation}
Using Eq.~(\ref{eq:rank-two-tensor})-(\ref{eq:optical-absorption}) we find the following compact expression for the optical absorption:
\begin{equation}\label{eq:absorption_tensor}
A_{ij}(\omega) = A_{\rm L}(\omega)\frac{v_{i}v_{j}}{v^2} + A_{\rm T}(\omega)\left(\delta_{ij} - \frac{v_{i}v_{j}}{v^2}\right)~.
\end{equation}
Equivalently, we can write Eq.~(\ref{eq:absorption_tensor}) more explicitly in terms of the angle $\psi$ in Eq.~(\ref{eq:defpsi}) between the polarization of the incident light and the direction of the current:
\begin{equation}\label{eq:final_absorption}
A_{ij}(\omega) = A_{\rm L}(\omega)\cos^{2}(\psi)  + A_{\rm T}(\omega)\sin^{2}(\psi)~.
\end{equation}
In the $T=0$ limit we have 
\begin{widetext}
\begin{eqnarray}\label{eq:longtransabsorption}
A_{{\rm L}({\rm T})}(\omega)&=& \pi \alpha_{\rm QED} \Theta(\hbar\omega - 2\mu)+ \alpha_{\rm QED} \Theta\left(\hbar\omega-\frac{2\mu}{1+ \beta}\right)\Theta\left(\frac{2\mu}{1 - \beta}-\hbar\omega\right)\nonumber\\
&\times&\left[\pm \frac{\hbar\omega-2\mu}{\beta\hbar\omega} \sqrt{1-\left(\frac{\hbar\omega-2\mu}{\beta\hbar\omega}\right)^{2}}-\text{sgn}\left(\frac{\hbar\omega-2\mu}{\beta\hbar\omega}\right)\arccos\left\vert\frac{\hbar\omega-2\mu}{\beta\hbar\omega}\right \vert \right]~.
\end{eqnarray}
\end{widetext}
Eqs.~(\ref{eq:final_absorption})-(\ref{eq:longtransabsorption}) are the most important results of this Section.
In writing these equations we have restored physical units for the sake of transparency. In the limit $\beta = 0$ Eq.~(\ref{eq:longtransabsorption}) yields $A_{\rm L}(\omega)= A_{\rm T}(\omega) = \pi \alpha_{\rm QED} \Theta(\hbar\omega - 2\mu)$. 

In Fig. \ref{Fig:AbsCurves} we show absorption as a function of photon energy $\hbar\omega$ (in units of $\mu$) for $\psi = 0$ (light polarization parallel to drift velocity) and $\pi/2$ (light polarization perpendicular to drift velocity) and different values of drift velocities (i.e.~of the ratio $\beta = v/v_{\rm F}$), at zero and finite temperature. We clearly see that the net effect of the finite drift velocity is to smooth the 
sharp absorption step $\pi\alpha_{\rm QED}\Theta(\hbar \omega- 2\mu)$ that occurs at $\beta =0$. The latter stems from Pauli blocking, namely the lack of available empty states in conduction band for vertical inter-band transitions assisted by photons with energy $\hbar \omega< 2 \mu$. Absorption is allowed only for $\hbar\omega > 2\mu$. Now, for a finite drift velocity $\beta$, a window of photon energies opens up between the $\beta$-dependent boundaries over which the chemical potential is tilted by the applied dc current:
\begin{equation}\label{eq:muplusminus}
\mu^{\pm } \equiv \frac{\mu}{1 \mp \beta}~,
\end{equation}
with $\mu^+>\mu$ and $\mu^-<\mu$. As illustrated in Fig.~\ref{Fig:Sketch}, it is possible to excite electron-hole pairs if the photon energy surpasses
the lower threshold $2\mu^{-}< 2\mu$. On the other hand, Pauli blocking extends its action {\it above} $2\mu$ in the window $2\mu<\hbar \omega<2\mu^+$.

\begin{figure}[tb]
\centering
\includegraphics[width= 8.5cm]{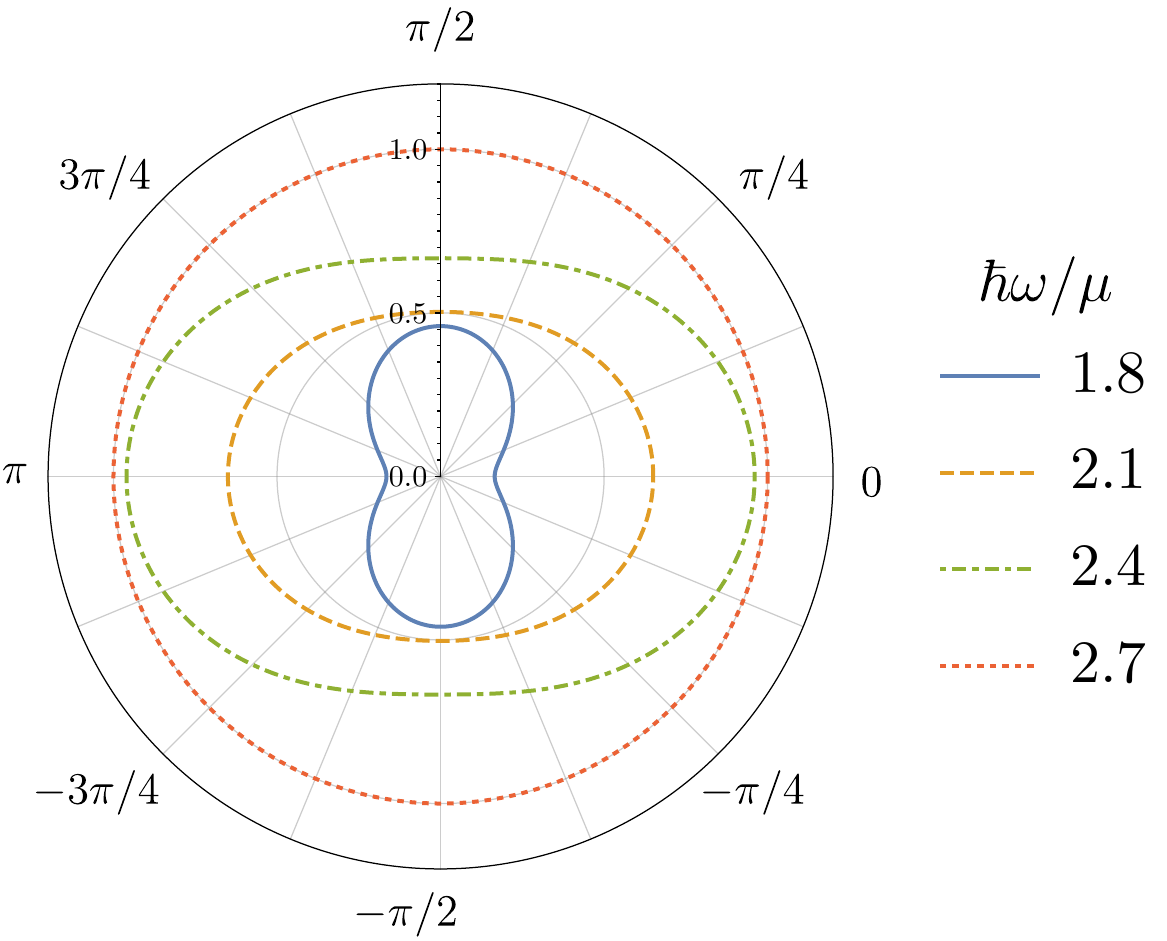}
\caption{(Color online) Angular dependence of the absorption $A_{ij}(\omega)$ (in units of the universal value $\pi\alpha_{\rm QED}$) as from Eq.~(\ref{eq:final_absorption}), for different values of the incident photon energy $\hbar \omega$ (in units of $\mu$), as indicated in the legend, and $\beta = 0.2$. In this plot we have set $T=0$.\label{Fig:PolarAbs}}
\end{figure}

In Fig. \ref{Fig:PolarAbs} we show that the way in which the 
sharp $\beta =0$ absorption step $\pi\alpha_{\rm QED}\Theta(\hbar \omega- 2\mu)$ 
is smoothed depends on the angle $\psi$ between in-plane light polarization and drift velocity ${\hat {\bm v}}$ in two major directions, $\psi =0$ and $\psi=\pi/2$. This results in a {\it birefringent} absorption in a current-carrying graphene sheet.

\begin{figure}[tb]
\centering
\includegraphics[width= 8.5cm]{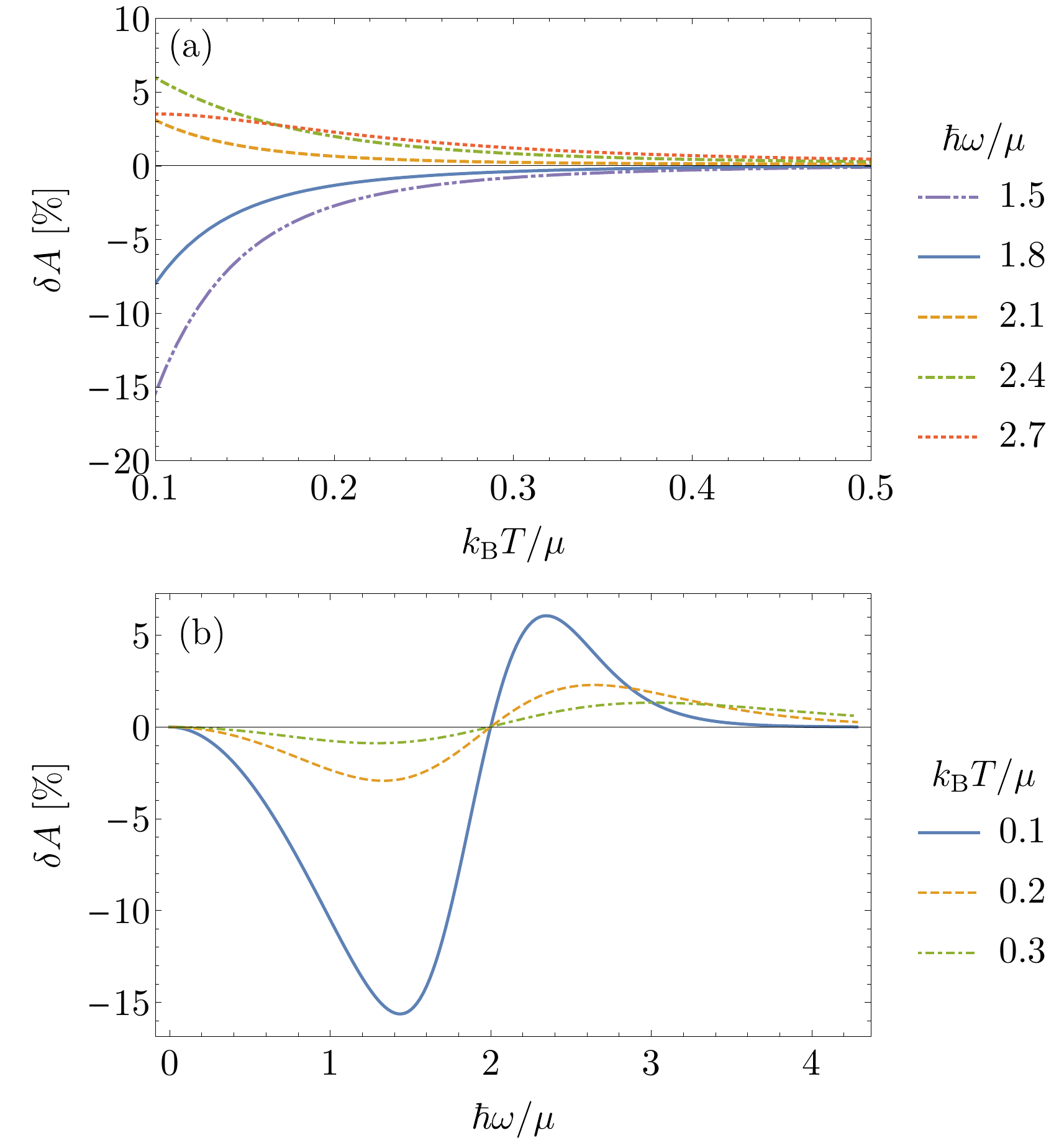}
\caption{(Color online) A measure of current-induced optical birefringence in a graphene sheet, as defined in Eq.~(\ref{eq:delta_A_anisotropy}). In panel (a) we show $\delta A$ as a function of temperature $T$ for different photon energies $\hbar \omega$ as indicated in the legend. In panel (b) we show $\delta A$ as a function of photon energy $\hbar \omega$, for different values of temperature $T$. \label{Fig:delta_A}}
\end{figure}

In order to quantify the degree of absorption anisotropy and its dependence on temperature, we calculate the quantity
\begin{equation}\label{eq:delta_A_anisotropy}
\delta A (\omega) = \frac{A_{\rm L}(\omega)-A_{\rm T}(\omega)}{A_{\rm L}(\omega)+A_{\rm T}(\omega)}~.
\end{equation}
Figs.~\ref{Fig:delta_A}(a) and (b) show the temperature and photon frequency dependence of $\delta A(\omega)$. Note that a finite temperature yields a degree of optical anisotropy that is spread over a wider spectral range, compared with the zero-temperature case. For example, Fig.~\ref{Fig:PolarAbs} shows that at $\hbar \omega = 2.7~\mu$ the absorption is isotropic at $T=0$, while Fig.~\ref{Fig:delta_A}(b) shows that the anisotropy $\delta A(\omega)$ is still significant at that photon frequency and finite temperature.

We emphasize that Figs.~\ref{Fig:AbsCurves} and \ref{Fig:delta_A} contain absorption data at finite temperature $T$, which have been obtained by using the Maldague approach described in Sect.~\ref{sect:finiteT}.
\section{Current-induced non-reciprocity and collimation of Dirac plasmons}
\label{sect:plasmons}

Having calculated the finite-temperature response function in Eq.~(\ref{eq:finite_temp_chi}), one finds the dielectric screening function 
in the random phase approximation~\cite{Giuliani_and_Vignale}:
\begin{equation}\label{eq:dielectric_function}
\varepsilon({\bm q},\omega;T) = 1- v_{{\bm q},\omega} \chi^{(0)}_{nn}({\bm q},\omega;T)~.
\end{equation}
Here $v_{{\bm q}, \omega} = 2\pi e^{2}/[\epsilon(\omega) q]$ is the 2D Fourier transform of the Coulomb potential, with $\epsilon(\omega)$ an effective frequency-dependent environmental dielectric constant and $-e<0$ the electron charge. For the case of graphene on hexagonal boron nitride (hBN)~\cite{principi_prb_2014,tomadin_prl_2015}, for example, 
$\epsilon(\omega) = [1 +\sqrt{\epsilon_{x}(\omega)\epsilon_{z}(\omega)}]/2$, while for the case of graphene encapsulated between two hBN slabs $\epsilon(\omega) = \sqrt{\epsilon_{x}(\omega)\epsilon_{z}(\omega)}$. In writing these equations we have neglected, for the sake of simplicity, finite-thickness effects~\cite{dai_science_2014,caldwell_naturecommun_2014,tomadin_prl_2015} (i.e.~Fabry-P\'erot phonon-polariton modes). 

Plasmons can be found as the roots of the real part of Eq.~(\ref{eq:dielectric_function}) on the real-frequency axis, when (Landau) damping is absent or small. Alternatively, they show up as strong absorption peaks in the {\it loss function}
\begin{equation}
L({\bm q},\omega;T) \equiv -{\rm Im}\left[\frac{1}{\varepsilon({\bm q},\omega;T)}\right]~,
\end{equation}
which is directly measured by electron-energy-loss spectroscopy~\cite{egerton_rpp_2009}. The loss function displays sharp peaks at the plasmon poles and carries also precious information on inter-band transitions and Landau damping.

\subsection{Non-reciprocity of plasmons}
\label{sect:non_reciprocity}

In the limit~$\bar{q}<\bar{\omega} \ll \mu$, the real part of the response function in Eq.~(\ref{PrincipalValue}) can be expanded as
\begin{eqnarray}\label{eq:real_part_approximation}
\frac{{\rm Re}[\chi^{(0)}_{nn}(\bm{q},\omega; 0)]}{D(\mu)} &\approx &    \frac{W(\beta)}{2\beta}\frac{\bar{q}^{2}}{\bar{\omega}^{2}}-\frac{\bar{q}^{2}}{8}+\gamma\frac{W(\beta)^{2}}{4\beta}\frac{\bar{q}^{3}}{\bar{\omega}^{3}}\nonumber \\
&+& \gamma \frac{\beta}{32}\frac{\bar{q}^{3}}{\bar{\omega}} + \frac{\bar{q}^{4}}{8\bar{\omega}} + \frac{3}{2}\frac{W(\beta)^{2}}{4\beta^{2}}\frac{\bar{q}^{4}}{\bar{\omega}^{4}}+\dots~,\nonumber\\
\end{eqnarray}
where $\gamma = +1~(-1)$ denotes ``upstream'' (``downstream'') plasmon propagation, i.e.~${\bm q} \cdot {\hat {\bm v}}/q = +1$ (${\bm q} \cdot {\hat {\bm v}}/q = -1$), and
\begin{equation}
\label{eq:W_equation}
W(\beta) = 2\frac{1-\sqrt{1-\beta^2}}{\beta}~.
\end{equation}
Note that $W(\beta) =\beta$ for $\beta \to 0$.  Eq.~(\ref{eq:real_part_approximation}) is correct up to ${\cal O}({\bar q}^4)$. 

Using this result, it is possible to find an approximate 
expression for the plasmon dispersion in the presence of current flow and in the long-wavelength ${\bar q}\ll 1$ limit. For the case of a frequency-independent dielectric constant $\epsilon$, we find
\begin{eqnarray}\label{eq:plasmon_dispersion}
\omega_{\rm pl}(q) &=& \sqrt{\frac{2 {\cal D}_0 W(\beta)}{\epsilon \beta}} \sqrt{q} \left[1 + \gamma \frac{\sqrt{2\beta W(\beta)}}{4} \sqrt{\frac{q}{k_{\rm TF }}} \right. \nonumber \\
&+& \left. \frac{12 - N^2_{\rm f}\alpha^2_{\rm ee}-3\beta W(\beta)}{16}\frac{q}{k_{\rm TF}} +\dots \right]~,
\end{eqnarray}
where ${\cal D}_0 = 4E_{\rm F} \sigma_{\rm uni}/\hbar$ is the Drude weight of non-interacting 2D MDFs~\cite{grigorenko_naturephoton_2012}, $k_{\rm TF} = N_{\rm f}\alpha_{\rm ee} k_{\rm F}$ is the $T=0$ Thomas-Fermi screening wave vector~\cite{kotov_rmp_2012}, and $\alpha_{\rm ee}=e^{2}/(\epsilon \hbar v_{\rm F})$ is graphene's fine structure constant~\cite{grigorenko_naturephoton_2012,kotov_rmp_2012}. 

\begin{figure}[tb]
\centering
\includegraphics[width= 8.5cm]{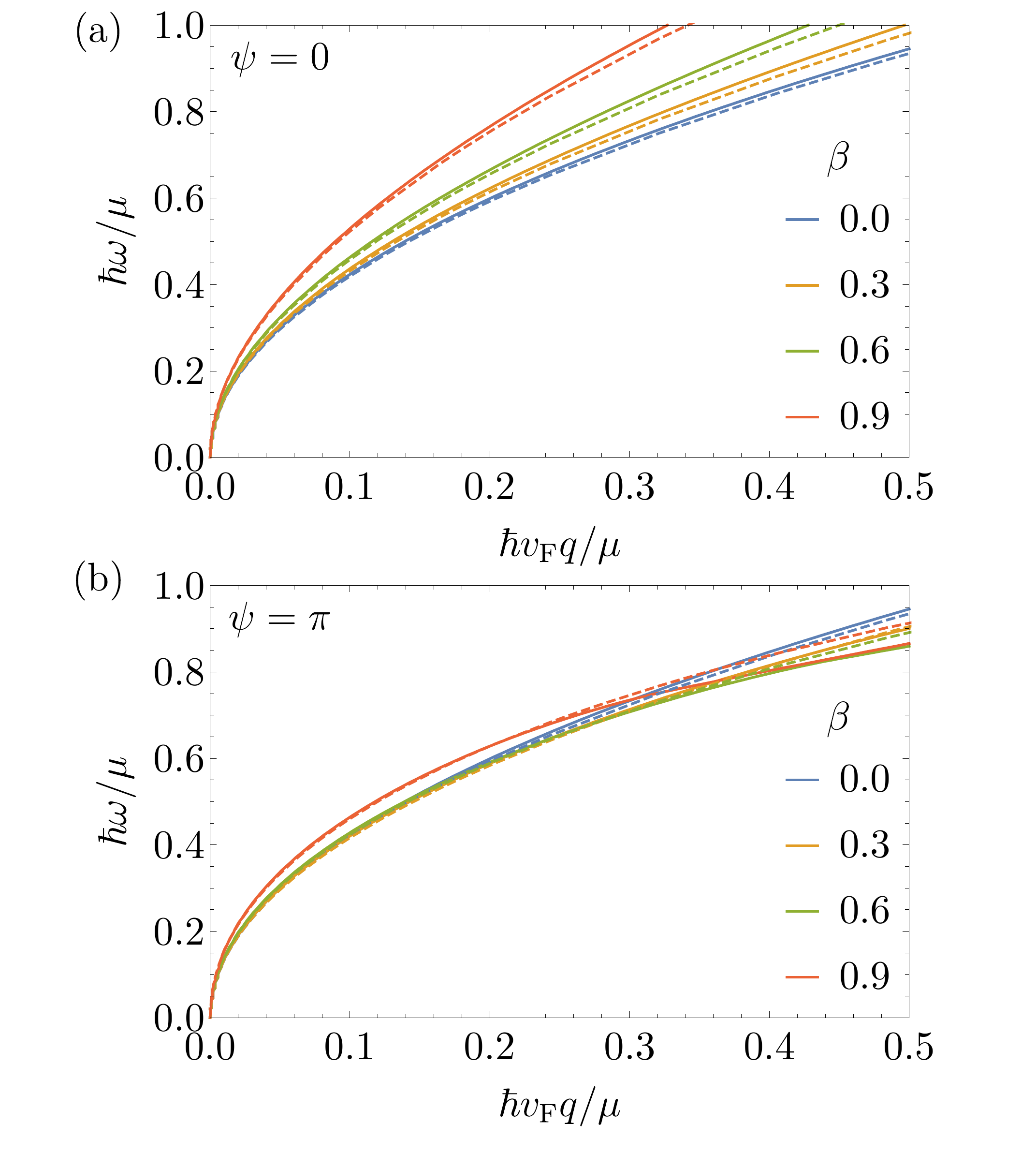}
\caption{(Color online) Dependence of the plasmon dispersion on the drift velocity $\beta$. Panel (a): upstream plasmons. Panel (b): downstream plasmons. Solid curves denote results obtained by numerically calculating the roots of Eq.~(\ref{eq:dielectric_function}). Dashed curves denote analytical long-wavelength results based on Eq.~(\ref{eq:plasmon_dispersion}). Results in this figure refer to $T=0$ and 
$\epsilon = 2.5$. With the scaling indicated in the horizontal and vertical axes they are ``universal", in that they do not depend on the chemical potential $\mu$.\label{Fig:non_reciprocity}}
\end{figure}

Eq.~(\ref{eq:plasmon_dispersion}), which is the most important result of this Section, shows that the dc current flow affects the Dirac plasmon dispersion in two ways. On the one hand, it renormalizes the Drude weight, i.e.~${\cal D}_0 \to {\cal D}_0 W(\beta)/\beta$.  On the other hand, it adds a linear-in-$q$ subleading term to the Dirac plasmon dispersion, whose sign depends on $\gamma = \pm 1$. Fig.~\ref{Fig:non_reciprocity} shows a comparison between the analytical result (\ref{eq:plasmon_dispersion}) (dashed lines) and the plasmon dispersion (solid lines) calculated numerically from the root of Eq.~(\ref{eq:dielectric_function}).
\subsection{Plasmon collimation}
\label{sect:collimation}

The particle-hole excitation spectrum of a current-carrying 2D MDF differs from that in the absence of a current, i.e.~for $\beta=0$. 
A geometrical analysis shows that the region where single-particle excitations cannot exist, i.e.~where~${\rm Im}[\chi^{(0)}_{nn}({\bm q},\omega;0)]=0$, is formed by the intersection of two cones in the $({\bm q}, \omega>0)$ space:
\begin{equation}\label{Eq:first_cone}
\hbar \omega = \hbar v_{\rm F} |{\bm q}|~,
\end{equation}
and
\begin{equation}\label{Eq:second_cone}
\frac{2\mu }{1-\beta^{2}} - \hbar\omega  = \left\vert\hbar v_{\rm F}{\bm q} - \frac{2\mu  }{1-\beta^{2}}{\bm \beta} \right\vert~,
\end{equation}
where ${\bm \beta} = \beta {\hat {\bm v}}$. Eq.~(\ref{Eq:first_cone}) defines the usual boundary between the inter- and intra-band electron-hole continua, which is a cone with its apex pointing down.  Eq.~(\ref{Eq:second_cone}) represents another cone with its apex pointing up and shifted from the origin of the $({\bm q}, \omega)$ plane by $2\mu {\bm \beta}/(1-\beta^{2})$ along the ${\hat {\bm v}}$ direction and 
$2\mu/(1-\beta^{2})$ along the $\omega$ axis.

\begin{figure*}[tb]
\centering
\includegraphics[width= 12cm]{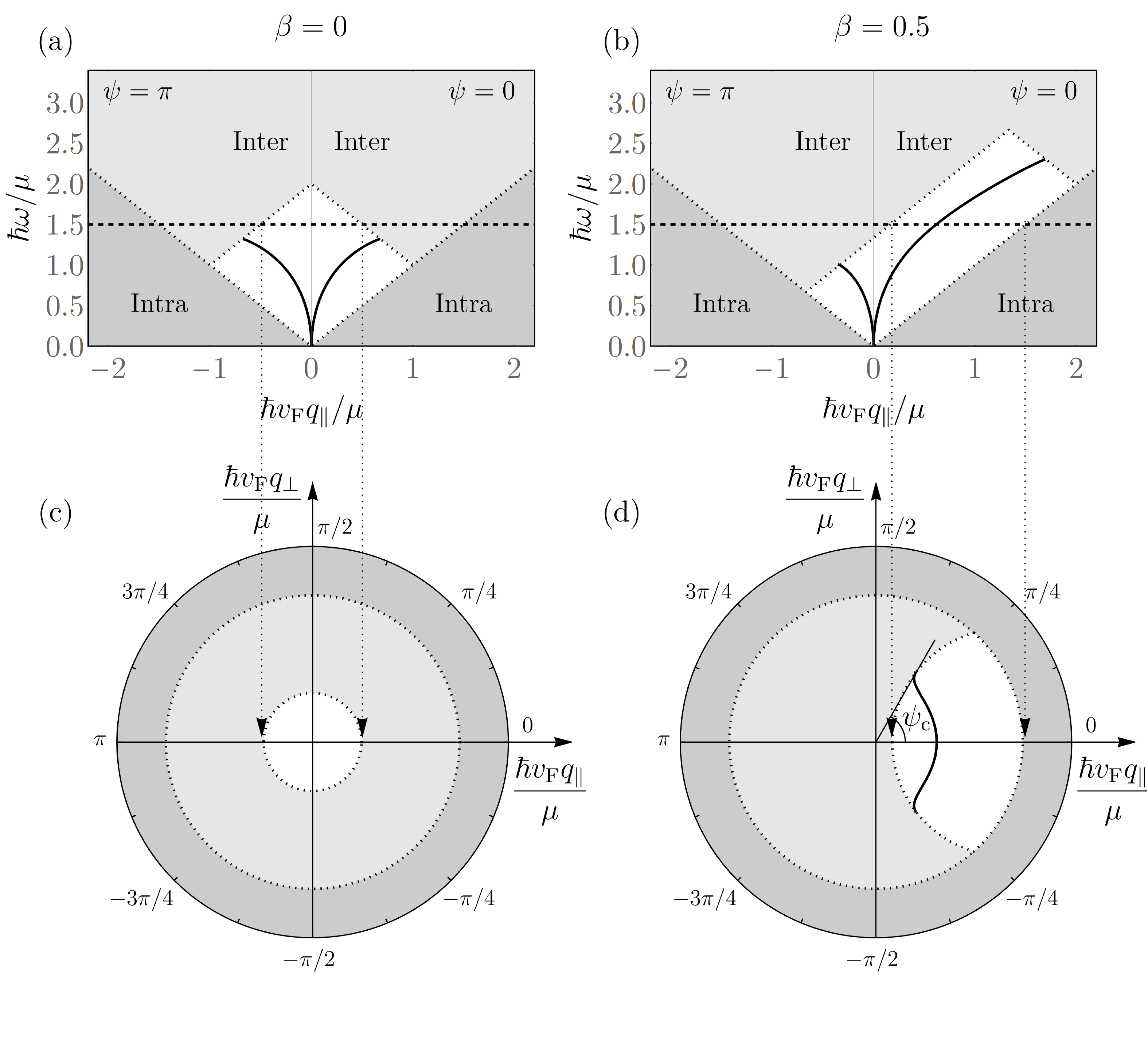}
\caption{Panels (a) and (b): Inter-band (light-grey shaded areas) and intra-band (dark-grey shaded areas) electron-hole continua in the $q_{\|}$-$\omega$ plane for $\beta = 0$ and $\beta = 0.5$, respectively. White areas indicate regions of parameters where no particle-hole excitations are possible at $T=0$, i.e.~where ${\rm Im}[\chi^{(0)}_{nn}({\bm q},\omega)]=0$. In these regions we have shown typical plasmon dispersions (solid lines) at $T=0$ and for $\epsilon(\omega)=1$. With the chosen units for $q_{\|}$ and $\omega$, the plasmon dispersion depends only on $\alpha_{\rm ee} = e^2/(\epsilon\hbar  v_{\rm F})$ (for a frequency-independent $\epsilon$) and $\beta$, but not on $\mu$. The dashed horizontal curve represents the frequency for which the plots in panels (c) and (d) have been made. In panel (a) we clearly see that, in the absence of an applied dc current, no plasmons can propagate at a photon frequency $\omega = 1.5~\mu/\hbar$ (since the plasmon dispersion hits the inter-band continuum at a lower value of $\omega$). Panels (c) and (d): Inter- and intra-band continua in the $(q_{\|}, q_{\perp})$ plane and for a photon frequency $\omega = 1.5~\mu/\hbar$. As explained above, in panel (c), for $\beta=0$, no plasmon branches fall in the region (white area) where ${\rm Im}[\chi^{(0)}_{nn}({\bm q},\omega)]=0$. On the contrary [panel (d)], we clearly see that an undamped collimated plasmon branch (solid line) emerges due to the application of a finite current. The quantity $\psi_{\rm c}$ denotes the collimation angle.\label{Fig:folding_regions_phes}} 
\end{figure*}

In Fig.~\ref{Fig:folding_regions_phes} we illustrate the relation between the $\beta =0$ and $\beta \neq 0$ inter- and intra-band continua. In particular, in Figs.~\ref{Fig:folding_regions_phes}(a) and~(b) we show the inter- and intra-band continua in the plane $(q_{\|},\omega>0)$ with $q_{\|} = q\cos{(\psi)}$, for $\psi =0$ and $\psi = \pi$. Figs.~\ref{Fig:folding_regions_phes}(c) and~(d) show the same quantities for a fixed value of $\omega>0$ and arbitrary $(q_{\|}, q_{\perp})$, where  $q_{\perp} = q\sin{(\psi)}$. The intersection between the two cones (\ref{Eq:first_cone}) and (\ref{Eq:second_cone}) yields the circular arcs in Fig.~\ref{Fig:folding_regions_phes}(d).

\begin{figure}[t]
\centering
\includegraphics[width= 8.5cm]{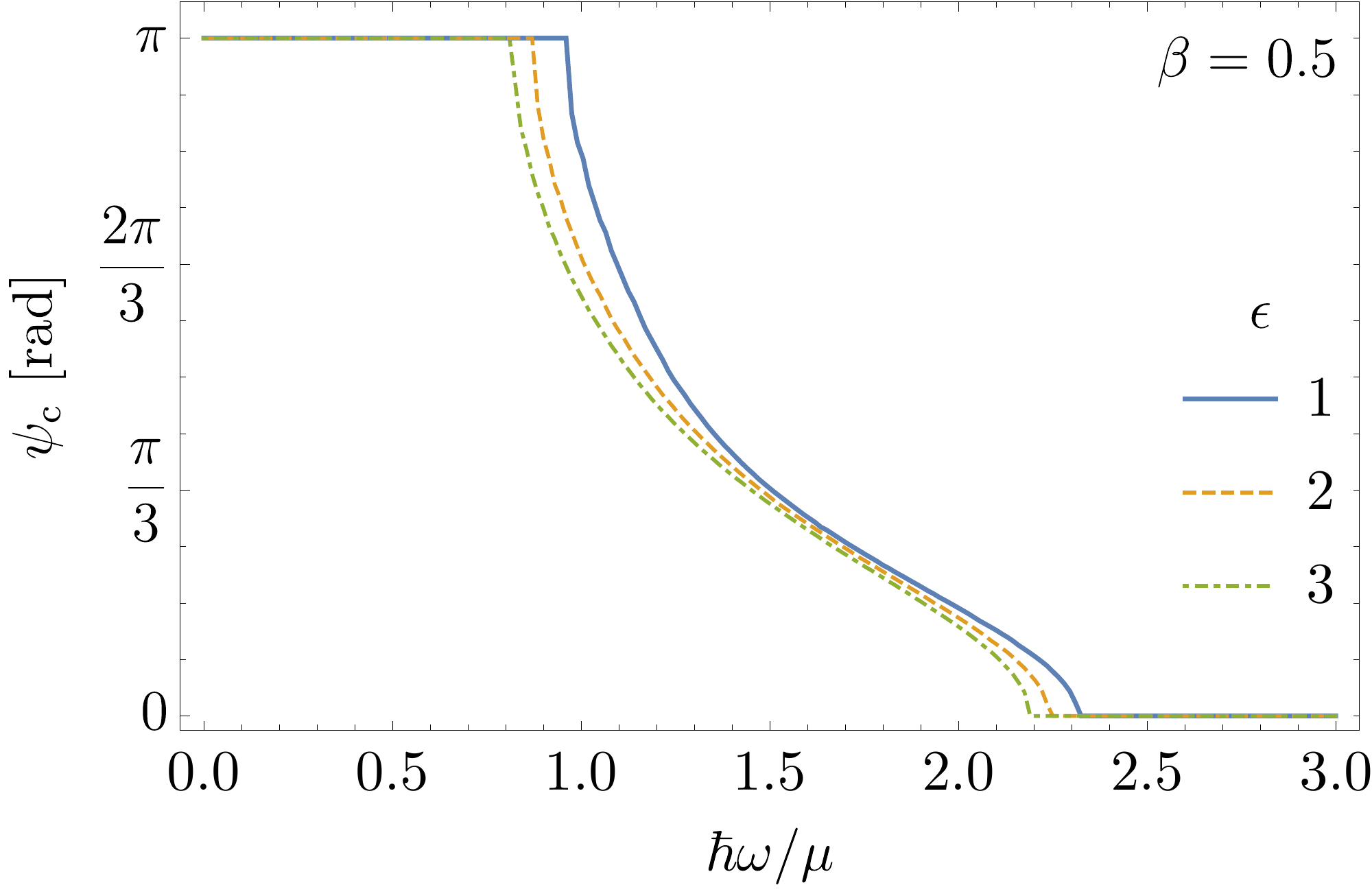}
\caption{(Color online) The collimation angle $\psi_{\rm c}$, measuring the directionality of the plasmon in a current-carrying graphene sheet, is plotted as a function of $\omega$ (in units of $\mu/\hbar$), for $\beta = 0.5$. The smaller the collimation angle, the stronger the directionality of plasmon propagation. On the other hand, $\psi_{\rm c} =\pi$ means non-collimated plasmons, i.e.~plasmons propagating in all directions. We clearly see that above a certain frequency threshold, undamped plasmon propagation at $T=0$ occurs in a range of angles $-\psi_{\rm c} \leq \psi \leq \psi_{\rm c}$.
When $\psi_{\rm c} = 0$ the plasmon lies entirely inside the inter-band continuum. Different curves refer to different values of the effective dielectric constant $\epsilon$, as indicated in the legend.\label{Fig:collimation_angle}} 
\end{figure}

In Fig.~\ref{Fig:folding_regions_phes} we also show plasmon modes obtained from the solution of the equation ${\rm Re}[\varepsilon({\bm q},\omega;T)]=0$ at $T=0$. For the sake of simplicity (to avoid the complications due to the hybridization between plasmons and the hBN optical phonon modes) we have set $\epsilon(\omega)=1$. We clearly see that a finite value of $\beta$ implies plasmon collimation in a window of angles $-\psi_{\rm c} \leq \psi \leq \psi_{\rm c}$ around the direction ${\hat {\bm v}}$ of current flow. This is because the lower edge of the inter-band continuum strongly depends on $\psi$: above a certain frequency threshold---see Fig.~\ref{Fig:collimation_angle}---only plasmon modes inside the window $-\psi_{\rm c} \leq \psi \leq \psi_{\rm c}$ cannot decay by emitting single electron-hole pairs.

\begin{figure*}[tb]
\centering
\includegraphics[width= 17cm]{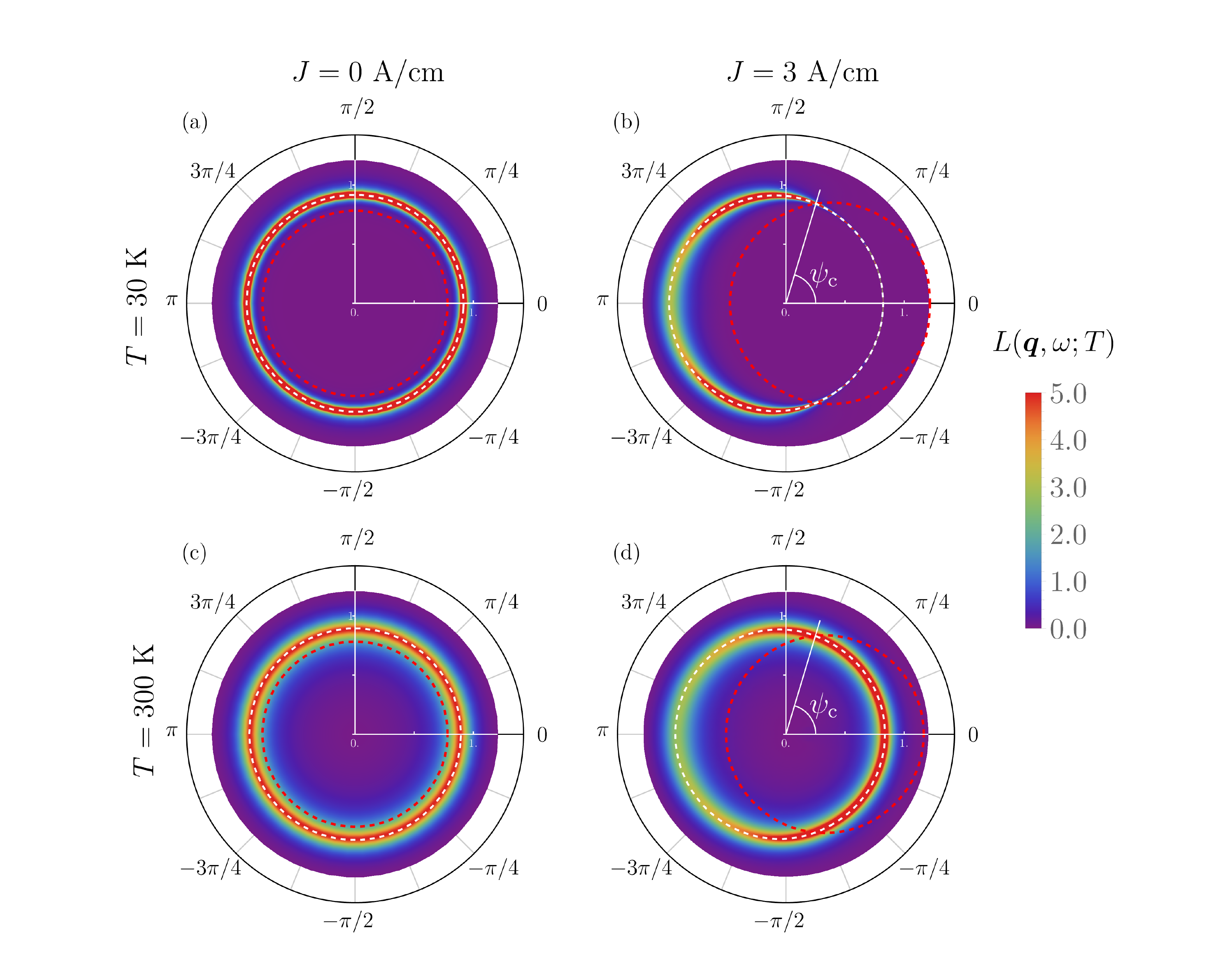}
\caption{(Color online) Energy loss function $L({\bm q},\omega; T)$ in the $(q_{\|}, q_{\perp})$ plane at an energy $\hbar \omega = 140~{\rm meV}$ and for a chemical potential $\mu = 115~{\rm meV}$. These results are for a graphene sheet on hBN, i.e.~for $\epsilon(\omega)=[1 +\sqrt{\epsilon_{x}(\omega)\epsilon_{z}(\omega)}]/2$. The regions inside the curves delimited by red dashed lines are regions of parameters in which there is no Landau damping at $T=0$ (white regions in Fig.~\ref{Fig:folding_regions_phes}). The white dashed curves represent the solution of the equation ${\rm Re}[\varepsilon({\bm q},\omega;T)]=0$. 
Sharp peaks in the loss function arise due to the existence of a plasmon mode.
Panels (a) and (c) refer to the case of zero dc current ($\beta=0$). Panels (b) and (d) refer to the case of an applied dc current density $J = 3~{\rm A}/{\rm cm}$: 
$\beta \simeq 0.18$ for panel (b) and $\beta \simeq 0.16$ for panel (d). The collimation angle $\psi_{\rm c}$ is also shown.\label{Fig:finite_T_plasmon}}
\end{figure*}

In Figs.~\ref{Fig:finite_T_plasmon}(a) and~(c) we show the loss function $L({\bm q},\omega; T)$ calculated for graphene on hBN at $\hbar \omega = 140~{\rm meV}$, for a chemical potential $\mu = 115~{\rm meV}$, and in the absence of a current. The two panels report results for different values of the electron temperature $T$. The role of an applied current is shown in Figs.~\ref{Fig:finite_T_plasmon}(b) and~(d), where we present results for the same parameters but for a finite current $J=3~{\rm A}/{\rm cm}$. Comparing Fig.~\ref{Fig:finite_T_plasmon}(a) with Fig.~\ref{Fig:finite_T_plasmon}(b), we clearly see that plasmons that are fully Landau damped at $T=30~{\rm K}$ in the absence of an applied current become sharply defined modes in a relatively narrow range of angles $-\psi_{\rm c} \leq \psi \leq \psi_{\rm c}$ in the presence of a current. A similar conclusion is also reached at $T=300~{\rm K}$. 

The collimation angle $\psi_{\rm c}$, which is defined at $T=0$ by the intersection of the solution of ${\rm Re}[\varepsilon({\bm q},\omega;T)]=0$ (at a fixed energy $\hbar \omega$) with the lower threshold of the inter-band electron-hole continuum, is plotted in Fig.~\ref{Fig:collimation_angle} as a function of $\hbar\omega/\mu$. The results shown in this figure have been obtained for the simplified case of a frequency-independent effective dielectric constant $\epsilon$. In the case of an applied current, plasmons propagating in the range of angles $-\psi_{\rm c} \leq \psi \leq \psi_{\rm c}$ around the direction of the applied current suffer no inter-band Landau damping. We therefore conclude that an applied dc current allows to collimate the propagation of Dirac plasmons.
\begin{figure}[t]
\centering
\includegraphics[width= 8.5cm]{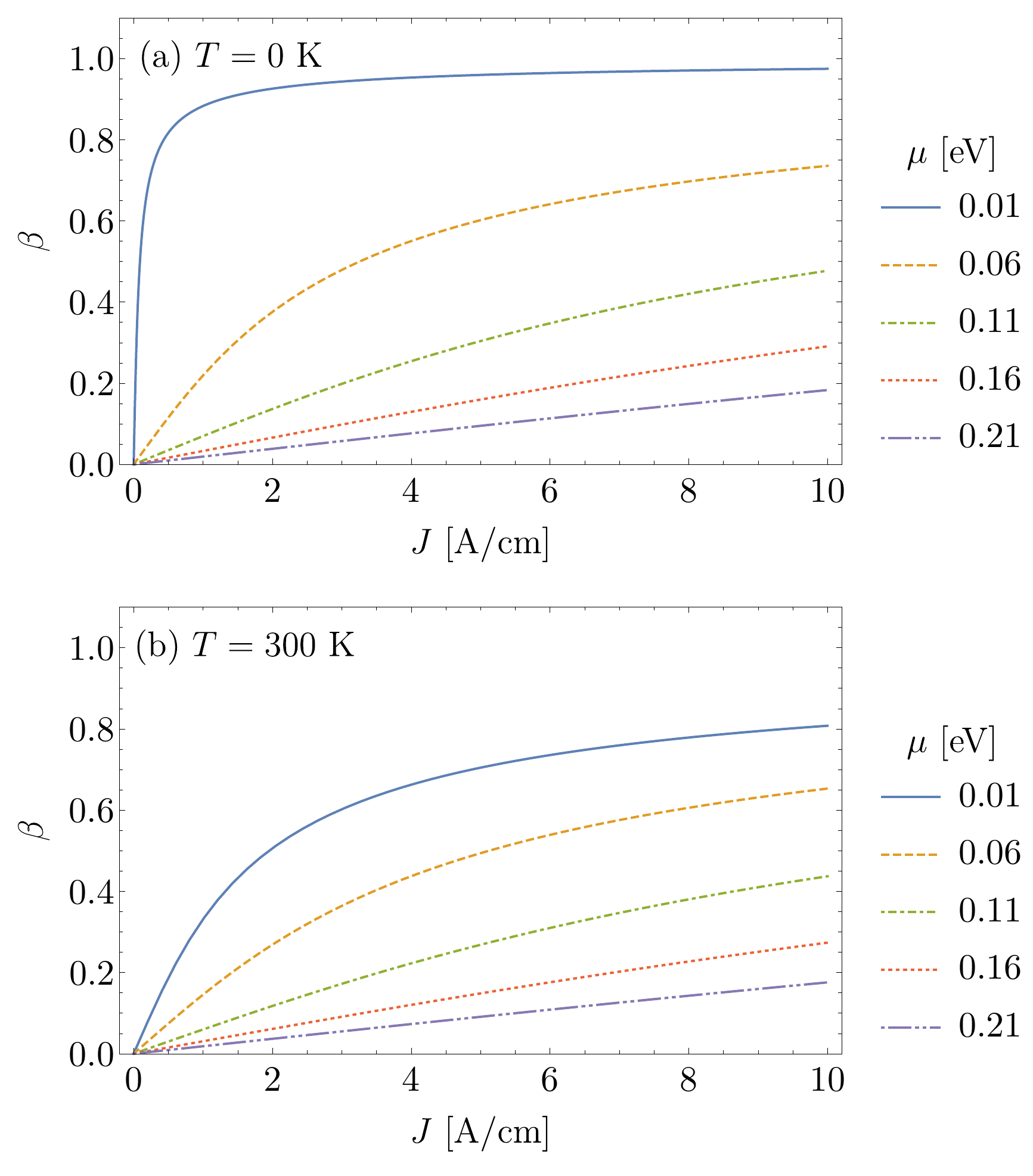}
\caption{(Color online) Dimensionless drift velocity $\beta$ as a function of the current density $J$ (in units of ${\rm A}/{\rm cm}$), for different values of the chemical potential $\mu$, as indicated in the legend. Panel (a): results at $T=0$. Panel (b): results at $T=300~{\rm K}$.\label{Fig:relation_j_v}}
\end{figure}
\section{Relation between $\mu$, $\beta$ and $n$, $J$}
\label{sect:relation}

Throughout this Article we have used the chemical potential $\mu$ and the dimensionless drift velocity $\beta$ as fundamental parameters. For practical purposes it may be useful to have results in terms of the carrier density $n_{\rm c}(\mu,T,\beta)$ and the current density $J(\mu,T,\beta)$. 
The latter quantities are obtained from $\mu$, $T$, and $\beta$ as
\begin{equation}\label{Eq:concentration}
n_{\rm c}(\mu,T,\beta) = \frac{n_{\rm c}(\mu,T )}{(1-\beta^{2})^{3/2}}
\end{equation}
and
\begin{equation}\label{Eq:CurrentDensity}
{\bm J}(\mu,T,\beta) = -e v_{\rm F} n_{\rm c}(\mu,T,\beta){\bm \beta}~,
\end{equation}
where $n_{\rm c}(\mu,T)$ is the carrier density in a graphene sheet at temperature $T$ and chemical potential $\mu$, in the absence of an applied dc current:
\begin{equation}\label{eq:carrier_density}
n_{\rm c}(\mu ,T) = n_{\rm e}(\mu ,T) + n_{\rm h}(\mu ,T)~.
\end{equation}
Here, $n_{\rm e}(\mu ,T)$ is the density of electrons in conduction band and $n_{\rm h}(\mu ,T)$ is the density of holes in valence band. 
In this Article we consider $\mu>0$ so holes are present only due to thermal excitation. The electron and hole densities in Eq.~(\ref{eq:carrier_density}) are obtained from the $\beta = 0$ Fermi-Dirac distribution function (\ref{eq:FDDistribution}):
\begin{eqnarray}
n_{\rm e}(\mu ,T) &=& S^{-1}\sum_{\bm k} n_{{\rm F},+} ({\bm k}, 0; T)\label{eq:electron_density}\\ 
n_{\rm h}(\mu ,T) &=& S^{-1}\sum_{\bm k} [1- n_{{\rm F},-} ({\bm k}, 0; T)]\label{eq:hole_density}~,
\end{eqnarray}
where $S$ is the sample area. The integrals in Eqs.~(\ref{eq:electron_density}) - (\ref{eq:hole_density}) can be written as
\begin{equation}\label{eq:carrier_density_integral}
n_{\alpha}(\mu ,T) = N_{\rm f} \int \frac{d^{2}{\bm k}}{(2 \pi)^2} M_{\alpha} (k, \mu ,T)~,
\end{equation}
where $\alpha = \pm 1$ corresponds to electrons (holes)  and
\begin{equation}
M_{\alpha} (k, \mu ,T) = \frac{1}{\exp{\left(\frac{\hbar v_{\rm F} k-\alpha \mu}{k_{\rm B}T}\right)} +1}~.
\end{equation}
The angular integration in Eq.~(\ref{eq:carrier_density_integral}) yields $2\pi$, while the radial integration gives:
\begin{equation}\label{eq:M_single_integral}
 \int_{0}^{\infty} \frac{d k}{2 \pi} kM_{\alpha} ( k, \mu ,T) = - \frac{(k_{\rm B}T)^{2}}{2\pi \left( \hbar v_{\rm F}\right)
^{2}} {\rm Li}_{2}\left[-e^{\alpha \mu /(k_{\rm B}T)}\right]~.
\end{equation} 
Here,  ${\rm Li}_{2}[x]$ is the dilogarithm function. This provides us with the well known expression for the carrier density in a graphene sheet
\begin{widetext}
\begin{equation}
n_{\rm c}(\mu ,T) = \frac{N_{\rm f} (k_{\rm B}T)^{2}}{2\pi \left( \hbar v_{\rm F}\right)
^{2}}\left\{-{\rm Li}_{2}\left[-e^{\mu /(k_{\rm B}T)}\right] - {\rm Li}_{2}\left[
-e^{-\mu /(k_{\rm B}T)}\right] \right\}~.
\end{equation}
\end{widetext}

To obtain Eq.~(\ref{Eq:concentration}) we need to calculate Eqs.~(\ref{eq:electron_density})-(\ref{eq:hole_density}) with the distribution function (\ref{eq:FDDistribution}). We find
\begin{equation}\label{eq:carrier_density_integral_drifted}
n_{\rm c}(\mu ,T, \beta) = N_{\rm f} \sum_{\alpha = \pm 1} \int \frac{d^{2}{\bm k}}{(2 \pi)^2} N_{\alpha} ({\bm k}, \mu ,T,\beta)~,
\end{equation}
where 
\begin{equation}\label{eq:M_function_drifted}
N_{\alpha} ({\bm k}, \mu ,T,\beta) = \frac{1}{\exp{\left\{\frac{\hbar v_{\rm F} k [1 - \alpha \beta \cos(\theta)] - \alpha \mu}{k_{\rm B}T}\right\}} +1}~.
\end{equation}
In Eq.~(\ref{eq:M_function_drifted}), $\theta$ is the angle between ${\bm k}$ and ${\bm v}$. Without loss of generality, the integral in Eq.~(\ref{eq:carrier_density_integral_drifted}) can be calculated by choosing ${\bm v}$ to be oriented along the ${\hat {\bm x}}$ direction. To calculate the integral one can replace the radial integration variable $k$ by introducing $k^{\prime} = k[1-\alpha \beta \cos(\theta)]$ such that the integral becomes
\begin{eqnarray}\label{eq:M_double_integral}
\int \frac{d^{2}{\bm k}}{(2 \pi)^2} N_{\alpha} ({\bm k}, \mu ,T,\beta) &=& \int_{0}^{2\pi} \frac{d\theta}{2\pi}\frac{1}{[1-\alpha \beta\cos{(\theta)}]^2}\nonumber \\
&\times & \int_0^{\infty}\frac{dk^\prime}{2\pi}k^\prime M_{\alpha}(k^{\prime},\mu,T)~.\nonumber\\
\end{eqnarray}
The radial integral in Eq.~(\ref{eq:M_double_integral}) is the same as in Eq.~(\ref{eq:M_single_integral}), while the angular integration can be carried out by using the residue theorem:  for $0<\beta<1$ we find
\begin{equation}\label{eq:angular_integral}
\int_{0}^{2\pi} \frac{d\theta}{2\pi}\frac{1}{[1-\alpha \beta\cos{(\theta)}]^2} =\frac{1}{(1-\alpha^2 \beta^2)^{3/2}}~.
\end{equation}
The desired result (\ref{Eq:concentration}) follows from Eqs.~(\ref{eq:M_double_integral})-(\ref{eq:angular_integral}).

To obtain Eq.~(\ref{Eq:CurrentDensity}) one proceeds in a similar way, by carrying out an integral over ${\bm k}$ of the expression for the current density in wave-vector space, i.e.
\begin{equation}\label{eq:current_density_integral_drifted}
{\bm J}(\mu ,T, \beta) =-e v_{\rm F} N_{\rm f} \sum_{\alpha = \pm 1} \int \frac{d^{2}{\bm k}}{(2 \pi)^2} \frac{\bm k}{k} \alpha N_{\alpha} ({\bm k}, \mu ,T,\beta)~,
\end{equation}
where the factor $\alpha$ describes the fact that holes contribute to the current density with the opposite sign with respect to electrons.
Similarly to Eq.~(\ref{eq:M_double_integral}), we can write
\begin{eqnarray}\label{eq:M_double_integral_current}
\int \frac{d^{2}{\bm k}}{(2 \pi)^2} \frac{\bm k}{k} N_{\alpha} ({\bm k}, \mu ,T,\beta) &=&{\hat {\bm v}} \int_{0}^{2\pi} \frac{d\theta}{2\pi}\frac{\alpha \cos(\theta)}{[1-\alpha \beta\cos{(\theta)}]^2}\nonumber \\
&\times & \int_0^{\infty}\frac{dk^\prime}{2\pi}k^\prime M_{\alpha}(k^{\prime},\mu,T)~.\nonumber\\
\end{eqnarray}
We then notice that
\begin{equation}\label{eq:partial_gbeta}
\int_{0}^{2\pi} \frac{d\theta}{2\pi}\frac{\alpha \cos{(\theta)}}{[1-\alpha \beta\cos{(\theta)}]^2} =  \frac{\partial g_\alpha(\beta)}{\partial \beta}
\end{equation}
with
\begin{equation}\label{eq:gbeta}
g_\alpha(\beta) = \int_{0}^{2\pi} \frac{d\theta}{2\pi}\frac{1}{1-\alpha \beta\cos{(\theta)}} = \frac{1}{\sqrt{1-\alpha^2 \beta^2}}~.
\end{equation}
Combining Eqs.~(\ref{eq:M_double_integral_current})-(\ref{eq:gbeta}) with Eq.~(\ref{eq:current_density_integral_drifted}) we finally find Eq.~(\ref{Eq:CurrentDensity}).

In Fig.~\ref{Fig:relation_j_v}(a) we show the dimensionless drift velocity $\beta$ as a function of the current density $J$, for different values of the chemical potential. We clearly see that for low current densities the relation is linear as expected from Eq.~(\ref{Eq:CurrentDensity}). With increasing current density the function $\beta=\beta(J)$ flattens, and $\beta\to 1$ (i.e. the induced drift velocity reaches the graphene Fermi velocity) asymptotically, as expected from Eq.~(\ref{Eq:CurrentDensity}). Finite-temperature effects, shown in Fig.~\ref{Fig:relation_j_v}(b), tend to suppress the induced drift velocity.

\section{Summary and discussion}
\label{sect:summary}

In this Article we have presented extensive calculations of the optical and plasmonic properties of a graphene sheet carrying a dc current. 

We have calculated in a fully analytical fashion the imaginary part of the density-density linear response function $\chi^{(0)}_{nn}({\bm q},\omega;T)$ of a current-carrying graphene sheet at zero temperature: our main results are reported in Sect.~\ref{sect:imag}, Eqs.~(\ref{Eq:SummaryImaginaryPart})-(\ref{eq:definitionsABC}) and Tables~\ref{Tab:Intra}-\ref{Tab:Inter}. The corresponding real part has been calculated by performing the one-dimensional quadrature in Eq.~(\ref{Eq:RealIntegral}). Results at finite temperature have been obtained by utilizing the Maldague identity in Eq.~(\ref{eq:finite_temp_chi}). We emphasize that the latter results are not of mere academic interest, since the electron temperature can be significantly raised above the lattice temperature under the application of a large dc current. 

This accurate knowledge of the density-density response function of a current-carrying graphene sheet has allowed us to calculate the optical absorption spectrum $A_{ij}(\omega)$, which displays a birefringent character with respect to the angle $\psi$ between the in-plane light polarization and current flow: see Eq.~(\ref{eq:longtransabsorption}) and Figs.~\ref{Fig:AbsCurves}-\ref{Fig:PolarAbs}. 

Finally, we have used the non-interacting density-density response function $\chi^{(0)}_{nn}({\bm q},\omega;T)$ to calculate the dielectric screening function in the random phase approximation, the plasmon dispersion, and the loss function of a current-carrying graphene sheet. We have showed that graphene plasmons acquire a degree of non-reciprocity under the influence of a dc current, see Eq.~(\ref{eq:plasmon_dispersion}), and that their propagation can be collimated in a window of angles around the current direction, see Figs.~\ref{Fig:folding_regions_phes}-\ref{Fig:collimation_angle}.

Before concluding, we would like to briefly discuss the possibility to test our predictions experimentally. To observe the discussed effects one needs to pass a large current through a graphene flake and measure polarized optical reflection or transmission through the sample. However, the large current passing through graphene would imply large Joule losses. Hence, an efficient thermal sink is needed in order to keep the temperature of graphene in equilibrium with the thermal bath and, therefore, constant. This implies that we cannot use the same set-up that was used previously to show that graphene membranes can host large current densities. We have to immobilize the graphene flake on an inert and thermally conductive substrate that acts as the thermal sink and need to perform the measurements in an inert gas atmosphere or vacuum environment. In addition, because low temperatures are beneficial for the observation of the effect, the usage of a suitable optical cryostat is needed. Currently, experiments to detect the current induced birefringence in graphene are under way.

\acknowledgments
This work was supported by the EC under the Graphene Flagship program (contract no.~CNECT-ICT-604391) and MIUR through the program ``Progetti Premiali 2012'' - Project ``ABNANOTECH''. B.V.D. wishes to thank the Scuola Normale Superiore (Pisa, Italy) for the kind hospitality while this work was carried out and Research Foundation Flanders (FWO-Vl) for a PhD Fellowship.

\end{document}